\documentclass[letterpaper,twocolumn,10pt]{article}
\usepackage{usenix-2020-09}

\usepackage{tikz}
\usepackage{amsmath}
\usepackage{amsmath,amsfonts}
\usepackage{algorithmic}
\usepackage{graphicx}
\usepackage{textcomp}
\usepackage{xcolor}
\usepackage{epsfig}

\usepackage{enumerate}

\usepackage{graphicx}
\usepackage{balance}
\usepackage{comment}
\usepackage{listings}
\usepackage{color,soul}
\usepackage{colortbl}
\usepackage{moreverb}
\usepackage{framed}
\usepackage{multirow}
\usepackage{pgfplots}
\usepackage{multirow}
\usepackage{rotating}
\usepackage{makecell}
\usepackage{pifont}
\usepackage{array}
\usepackage{subfig}

\usepackage{tcolorbox}
\usepackage{cleveref}

\usepackage{tikz}

\makeatletter
\newif\if@restonecol
\makeatother

\usepackage[ruled,vlined,linesnumbered]{algorithm2e}
\definecolor{lightgray}{gray}{0.9}
\definecolor{lightblue}{rgb}{0.9,0.9,1}
\definecolor{red}{rgb}{1,0,0}

\let\ls\lstinline

\newcommand\cut[1]{}

\newcommand\candidate[1]{}

\newcolumntype{L}[1]{>{\raggedright\let\newline\\\arraybackslash\hspace{0pt}}m{#1}}

\usepackage{xspace}

\newcommand\os{operating system\xspace}

\newcommand\Sname{\texttt{EXACT}\xspace}

\definecolor{bella@base03}{HTML}{002B36}
\definecolor{bella@base02}{HTML}{073642}
\definecolor{bella@base01}{HTML}{586e75}
\definecolor{bella@base00}{HTML}{657b83}
\definecolor{bella@base0}{HTML}{839496}
\definecolor{bella@base1}{HTML}{93a1a1}
\definecolor{bella@base2}{HTML}{EEE8D5}
\definecolor{bella@base3}{HTML}{FDF6E3}
\definecolor{bella@yellow}{HTML}{B58900}
\definecolor{bella@orange}{HTML}{CB4B16}
\definecolor{bella@red}{HTML}{DC322F}
\definecolor{bella@magenta}{HTML}{D33682}
\definecolor{bella@violet}{HTML}{6C71C4}
\definecolor{bella@blue}{HTML}{268BD2}
\definecolor{bella@cyan}{HTML}{2AA198}
\definecolor{bella@green}{HTML}{859900}

\usetikzlibrary{tikzmark}

\begin{document}

\date{}

\title{\Large \bf Artificial-Intelligence Generated Code Considered Harmful: A Road Map for Secure and High-Quality Code Generation}

\author{
{\rm Chun Jie Chong, Zhihao (Zephyr) Yao, Iulian Neamtiu}\\
\{cc255, zhihao.yao, ineamtiu@njit.edu\}\\
New Jersey Institute of Technology\\
}

\maketitle

\begin{abstract}
Generating code via a LLM (rather than writing code from scratch),
has exploded in popularity.  However, the security implications of
LLM-generated code are still unknown.  We performed a study that
compared the security and quality of human-written code with that of
LLM-generated code, for a wide range of programming tasks, including
data structures, algorithms, cryptographic routines, and LeetCode questions.
To assess code security we used unit testing, fuzzing,
and static analysis. For code quality, we focused on complexity and size.
We found that LLM can generate incorrect code that fails to implement the required
functionality, especially for more complicated tasks; such errors can
be subtle. For example, for the cryptographic algorithm SHA1, LLM
generated an incorrect implementation that nevertheless compiles.
In cases where its functionality was correct, we found that LLM-generated code is less
secure, primarily due to the lack of defensive programming constructs,
which invites a host of security issues such as buffer overflows or integer overflows.
Fuzzing has revealed that LLM-generated code is more prone to hangs and crashes than
human-written code.  Quality-wise, we found that LLM generates bare-bones code
that lacks defensive programming constructs, and is typically
more complex (per line of code) compared to human-written code.  Next,
we constructed a feedback loop that asked the LLM to re-generate the
code and eliminate the found issues (e.g., malloc overflow, array
index out of bounds, null dereferences). We found that the LLM fails
to eliminate such issues consistently: while succeeding in some cases,
we found instances where the re-generated, supposedly more secure code,
contains new issues; we also found that upon prompting, LLM can
introduce issues in files that were issues-free before prompting.
Our study exposes the perils of LLM-generated code (and feedback loops),
particularly in the critical security domain.
\end{abstract}

\maketitle

\section{Introduction}
\label{sec:intro}
Software security is of utmost importance in software engineering, as it directly affects the security and reliability of a digital society increasingly dependent on software.
For example, the 2024 CrowdStrike bug that crashed 8.5 million Microsoft Windows devices~\cite{crowdstrike_bug}
highlights the worldwide impact of software bugs.
Human experts have been trained to write, review, and test code to ensure its quality, despite the fact that the process is time-consuming and error-prone.
However, the security and quality of Artificial Intelligence (AI)-generated code is an under-studied area.
With the advance of AI, the shift towards AI-assisted programming is rapidly gaining momentum,
making the concerns more urgent.

Large Language Models (LLMs) are already widely used to assist developers in code completion and summarization
~\cite{amazon_code_whisper, github_copilot, gemini_code_assist}, and in some cases, to automatically generate code from scratch to meet the requirements of specific tasks~\cite{romera2024mathematical}.
For example, GitHub Copilot~\cite{github_copilot} (``Copilot'' for short), a widely-adopted AI coding assistant, has been available since 2022~\cite{github_copilot}.
While users report an improvement in productivity (81\% and 88\%, respectively reported by two independent user studies~\cite{akvelon_copilot_study, github_productivity}), an empirical investigation has shown that 40\% of Copilot-generated programs are buggy~\cite{pearce2022asleep}.
The false sense of productivity when using LLM code generation in the workplace has mostly been driven by developers aiming to fulfill industry's internal performance metrics, such as task completion time and lines of code produced~\cite{github_study_ai_tool}, rather than code quality.

The
security implications of AI code generation are largely ignored by the industry:
Copilot Voice has been introduced as
a new feature that allows inexperienced developers to generate full programs by simply speaking to the AI assistant,
marketed as a new way to ``write code without the keyboard''~\cite{copilot_voice}.
Na\"ive trust in AI code generation can lead to significant security vulnerabilities and degradation in code quality.
Understanding the traits and limitations of the LLM-generated code is important to
facilitate the adoption of LLMs in future software engineering practice.
In October 2023, the White House issued an Executive Order on ``Safe, Secure, and Trustworthy Development and Use of Artificial Intelligence'', which emphasize the importance of security in such AI use cases~\cite{nist_ai_risk, ai_executive_order}.

Motivated by the national goal,
this work aims to study the security and quality of code generated by the state-of-the-art LLM model, OpenAI's GPT-4o, which powers GitHub Copilot Enterprise~\cite{github_copilot_gpt4o}.
Specifically, we create an evaluation framework, \Sname (
\underline{EX}amination System for \underline{A}I-Generated \underline{C}ode \underline{T}esting), which
uses a suite of program analyses, unit testing, fuzzing, complexity and size measurements, to assess code security and quality.
\Sname compares these factors with the human-written code that implements the same functionality.
We focus on three classes of representative coding tasks: over 200 LeetCode programming tests, fifteen commonly-used algorithms and data structure implementations, as well as popular cryptographic functions.
For each of these tasks, we compare the security and quality of the human-written and LLM-generated code.
We chose the C programming language due to its prevalence in systems and security-sensitive code.

For each comparison experiment,
\Sname checks functionalities
of LLM-generated and human-written code using an empirical method: if a predefined test suite (e.g., LeetCode online submission \cite{leetcode}) is available, we submit both programs to the test suite;
otherwise, we use unit testing and AFL~\cite{afl} to create test cases and compare the runtime behavior of the two programs.
Note that fuzzing findings are used to assess both security and functionality.
A recent study \cite{zhang2024cybench} demonstrates that LLM is not capable of solving complicated tasks (tasks that take a human more than 11 minutes). We observe a similar trait in our study: among the 21 LeetCode tasks that LLM failed to solve, 20 tasks are rated at medium or hard difficulty.
LLM also implements the SHA-1 hash function incorrectly, where it {\em generates the wrong hash values for all the inputs}.
We also demonstrate the unreliability of both the AI-generated and human-written code in the presence of fuzzing, as they both fail to handle corner cases in our test suites (though human code performs slightly better).

We also found that the security and quality of LLM-generated code are lacking, compared to its human-written equivalent.
The security issues found by Clang static analyzer in LLM-generated code are
consistently higher than human-written code: by 11.2\% in LeetCode and 7.1\% in algorithm tasks.
Additionally, LLM-generated code is 1.19x
more complex than human-written code in LeetCode and 1.26x in algorithm tasks.

A feedback loop is a state-of-the-art practice carried out by various research groups
\cite{kaelbling1996reinforcement,sutton2018reinforcement,mcaleese2024llm}
to improve LLM's code generation.
Therefore, our study also investigates the effectiveness of a feedback loop in improving the security and correctness of the generated code
by iteratively feeding the results of the evaluation back to the LLM model.
We show that when we asked the model to fix the discovered security issues, {\em it can introduce new bugs in code that was previously bug-free}; moreover, the re-generated version has higher complexity per line.
We also observe that by requesting shorter line lengths, GPT-4o yields subsequent code with 11.3\% lower cyclomatic complexity per line, and surprisingly, it also fixed 17.9\% of the
security issues found by Clang static analyzer.
In contrast,
by requesting fewer lines of code, GPT-4o yields code with 18.4\% higher complexity per line, and fixed 43.2\% of security issues.

During the course of our study, we have identified 116 security issues in public GitHub repositories, which we have responsibly disclosed to the repository owners.
For the software security issues of the code generated by OpenAI's GPT-4o model,
we have reached out to OpenAI for their further improvement.
We will open source \Sname and our test data.

The main contributions of our study are:
\begin{itemize}
\item We show that, for the same task, LLM generates less-secure and lower-quality code than the human equivalent
\item We characterize the substantial security issues in LLM-generated code
\item We demonstrate that prompt engineering with a feedback loop does not necessarily improve the security of LLM-generated code.
\end{itemize}

\section{Background}
\label{sec:bg}

\subsection{AI Code Generation}

LLMs have been widely adopted for natural language tasks, and also for code generation.
This adoption is easily justifiable as LLMs have shown exceptional performance in code completion~\cite{raychev2014code}, translation~\cite{yin2024rectifier}, and full project code generation~\cite{liventsev2023fully}.
With the introduction of large-scale pre-trained LLMs, such as OpenAI's GPT-4o that powers GitHub Copilot Enterprise~\cite{github_copilot_gpt4o}, the performance of AI code generation has been further advanced to a degree that it is presumed to replace human programmers~\cite{github_ceo}.

The inception of AI code generation has brought about both a revolution and a debate in the software community.
On one hand, AI code improves developers' productivity by automating coding tasks, especially the repetitive and mundane ones~\cite{dilhara2024unprecedented}.
However, the security implications of fully autonomous programming using LLMs~\cite{liventsev2023fully}
and automation of repetitive coding tasks~\cite{dilhara2024unprecedented}
that have been proposed by recent research are not well understood.

In a Copilot study, 88\% of human developers have reported that they are more productive when coding with the tool~\cite{github_productivity}.
The drastic improvement in productivity
demonstrates the potential of AI in software engineering.
Indeed, the CEO of GitHub
predicts that ``sooner than later,'' ``Copilot will write 80\% of code''~\cite{github_ceo}.
But,
unfortunately,
existing AI coding assistants have been shown to write incompetent code~\cite{pearce2022asleep, perry2023users, wang2024investigating}.
According to a study conducted by GitHub, 92\% of the surveyed developers have used AI coding tools, and 70\% of them believe that the tools offered an advantage in their work~\cite{github_study_ai_tool}.
One impetus for AI code generation is to improve companies' internal performance metrics, such as ``time to complete a task'', or ``lines of code written''~\cite{github_study_ai_tool}.
The adoption of AI-generated code has raised both ethics and quality concerns.

Recent studies have shown significant degradation in code quality~\cite{perry2023users, pan2024lost, pearce2022asleep}.
NetBSD, a popular open-source \os project, referred to AI-generated code as ``tainted code'' and banned it from their codebase~\cite{netbsd_ban_llm}.
We present an illustration of GPT-4o-generated code buffer size in \Cref{fig:memory_bug_background} as an example of the quality concerns.
We discuss this example in depth in \Cref{sec:motivation_preliminary}.

\begin{figure}[t]
\centering
\begin{minipage}{.45\textwidth}
\begin{lstlisting}[{xleftmargin=1em}]
void setBuffer(char *buffer) {
    for (int i = 0; i < @\tikzmark{circle_start}@26@\tikzmark{circle_end}@; i++)
        buffer[i] = '\0';
}

int main() {
    char alphabet[] = "ABCDEF@\tikzmark{missingG_start}@H@\tikzmark{missingG_end}@IJKLMNOPQRSTUVWXYZ";
    int length = 0;
    while (alphabet[length] != '\0')
        length++;
    char *buffer =
        (char *)malloc(length*sizeof(char));
    setBuffer(buffer);
    free(buffer);
    return 0;
}
\end{lstlisting}
\end{minipage}
\caption{Sample task 1 (\S\ref{sec:motivation_preliminary}): we prompt GPT-4o to find the size of the buffer on line~7 and use it on line~2 for the loop condition
(in the red circle). GPT-4o produces an incorrect answer, leading to heap overflow.}
\label{fig:memory_bug_background}
\end{figure}

\subsection{Automated Code Improvement}
\label{sec:ar_transformer}

LLMs have been using Reinforcement learning (RL) for automated code improvement.
RL updates a model's parameters through interactions with human feedback or the environment, and has been shown to be effective in improving various AI models' performance~\cite{kaelbling1996reinforcement, sutton2018reinforcement}.
Reinforcement learning from human feedback (RLHF) uses human feedback to fine-tune a model's parameters~\cite{christiano2017deep}.
Despite the fact that RLHF can be used to improve the performance of LLM code generation, scalable deployment in practice has been limited by the capability of human evaluators to provide meaningful feedback to the outputs~\cite{mcaleese2024llm}.
To cope with the lack of human feedback, OpenAI has proposed a fine-tuned model based on GPT 4, namely CriticGPT, to automatically
critique the
LLM-generated code, where the feedback is used to improve the subsequent generations
~\cite{mcaleese2024llm}.
A pool of bugs and human feedback originated from OpenAI's previous RLHF pipeline is used as the training set for the fine-tuning of CriticGPT~\cite{mcaleese2024llm}.
Likewise, GitHub Copilot has proposed a secondary LLM model to ``approximate the behavior of static analysis tools'' to
provide feedback to the outputs of the primary LLM model
~\cite{github_ai_bug_catching}.
Since these systems have been integrated into the LLM pipeline, our work aims to investigate the security of the overall system that may include these blackbox components.

\begin{tikzpicture}[remember picture, overlay]
\draw[red, thick] ([yshift=2pt, xshift=4.7pt]pic cs:circle_start) circle (5.5pt);
\draw[red, thick, <-] ([yshift=-3pt, xshift=9pt]pic cs:circle_start) -- ++(18pt, -8pt) node[below] {\textbf{GPT-4o fills in an incorrect value.}};
\draw[blue, thick, <-] ([yshift=5.6pt, xshift=0.8pt]pic cs:missingG_start) -- ++(5.6pt, 2pt) node[above] {\textbf{Intentionally missed letter `G'.}};
\end{tikzpicture}

\subsection{Implications of AI Code Generation}

Na\"ive trust in AI code-generation tools can lead to deteriorated code quality and software security hazards.
Due to the lack of metrics to evaluate the trustworthiness of generated code~\cite{wang2024investigating}, existing quantitative evaluation of AI-generated code is limited.
A user study conducted in 2023 has found significant degradation in code quality and a false sense of security when programmers use AI assistants~\cite{perry2023users}.
Likewise, Pearce et al. conducted an empirical study on 1,689 Copilot-generated programs and found that
40\% of the programs are vulnerable~\cite{pearce2022asleep}.
Another user study in 2024 has shown that
a majority of GitHub Copilot users felt that the tool's suggestions were not always accurate, as the tool ``may give false code suggestions that mislead developers''~\cite{akvelon_copilot_study}.
Fixing the errors in the generated code is a challenging task for most users because they did not author the code themselves. Consequently, when AI-generated code appears to be complex from the user's perspective, users often give up on fixing the code and resort to other online resources~\cite{vaithilingam2022expectation}.
Indeed, stringent testing shows that, currently, even the most advanced LLM models, such as Claude 3.5 Sonnet~\cite{claude_3_5_sonnet} and GPT-4o~\cite{openai_gpt4o}, are only able to solve basic, straightforward tasks, that take human developers at most 11 minutes to solve~\cite{zhang2024cybench}.

\section{Motivation}
\label{sec:motivation}

\subsection{Sample Task 1: Finding a Buffer's Size}
\label{sec:motivation_preliminary}

As motivation for our study, we present a preliminary experiment we conducted on OpenAI GPT-4o, where we found that the model
fails a basic, yet critical, security task: {\em finding the size of a buffer}, even though the buffer size is given in the prompt in a way that a human can easily understand.
The incorrect buffer size can lead to buffer overflow if it is larger, or leaking non-initialized memory if the buffer size is smaller than the actual buffer size in the buffer initialization function.
We present an example of wrong buffer size in \Cref{fig:memory_bug_background}, where the buffer size is filled in as 26, but the actual buffer size should be 25 (as the length of an incomplete alphabet with the letter `G' missing is 25).
In our experiment, we asked GPT-4o to fill in the buffer size in the loop condition (the red circle in \Cref{fig:memory_bug_background}) with various buffer size definitions in the referenced code in our prompt.
\Cref{tab:prelim_study_buffer_size} shows the buffer size definitions, examples, and the success rate of GPT-4o in filling in the correct buffer size.
We repeat experiments for each buffer size definition 1000 times. We found that GPT-4o not only fails to fill in the correct buffer size when calculation is involved, but also fails to fill in the constant buffer size in some cases.

\begin{table*}[h!]
\centering
\begin{tabular}{|l|c|c|}
\hline
\textbf{Buffer Size Definitions} & \textbf{Example} & \textbf{Success Rate}\\
\hline
A constant integer & 402 & 99.3\% \\
\hline
The length of a string (alphabet) & len(ABC...XYZ) & 92.8\% \\
\hline
The exponentiation of an integer to the power of another integer & pow(4, 2) & 91.4\% \\
\hline
Multiplying two integers & 415 * 495 & 63.1\% \\
\hline
Subtracting a float from an integer & 922 - 174.05 & 50.1\% \\
\hline
The exponentiation of an integer to the power of another integer, plus a float & pow(1,5) + 63.56 & 42.4\% \\
\hline
The length of a string (alphabet with a missing letter) & len(ABC...XYZ) & 41.6\% \\
\hline
A constant integer, but with irrelevant presence of length of a string & 612 & 31.2\%\\
\hline
Multiplying an integer and a float & 893 * 518.33 & 10.6\% \\
\hline
Multiplying two floats & 285.63 * 62.72 & 1.5\% \\
\hline
\end{tabular}
\caption{Preliminary study on GPT-4o's ability to correctly determine the buffer size in \Cref{fig:memory_bug_background}.}
\label{tab:prelim_study_buffer_size}
\end{table*}

\begin{figure}[h!]
\centering
\includegraphics[width=0.5\textwidth]{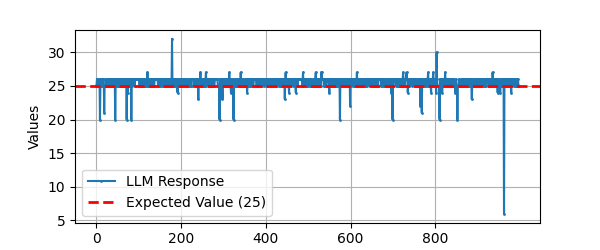}
\caption{GPT-4o filled buffer size (the red circle in \Cref{fig:memory_bug_background}) over 1000 trials.}
\label{fig:motivating_example_fig}
\end{figure}

As shown in \Cref{tab:prelim_study_buffer_size},
GPT-4o has a high success rate (but not 100\%) in filling in the buffer size when the buffer size is a constant integer. When the buffer size requires calculation, the success rate drops significantly.
The success rate is the lowest when the buffer size is calculated using floating-point numbers, possibly due to the model's lack of understanding of float to integer casting.
To our surprise, GPT-4o's success rate drops from 92.8\% to 41.6\% when the buffer size is calculated from the length of alphabet with a missing letter, showing that the model concludes the buffer size based on the pattern of the alphabet, rather than the actual length of the string.
\Cref{fig:motivating_example_fig} shows the distribution of the buffer size filled in by GPT-4o in the missing letter example over 1000 trials.
We were also surprised that when we give the model a constant integer buffer size, but at the same time,
present an irrelevant calculation of the length of a string, the success rate drops from 99.3\% to 31.2\%.
In rare cases (0.08\% of all results), the model has even filled the size (an expected \ls|int| value) with meaningless non-English characters and symbols.
This preliminary study demonstrates that GPT-4o has a high chance of generating memory bugs in C code, and motivates us to further investigate the security of LLM-generated code.

\subsection{Sample Task 2: Implementing SHA1}
\label{sec:sha1_sample}

We asked GPT-4o to complete a widely-used, security-critical algorithm: SHA1 (Secure Hash Algorithm 1). The SHA1 header file from OpenBSD's GitHub repository~\cite{openbsd} was provided in the prompt to make sure that the specifications are clear, and there are no discrepancies in function input/output formats
between the GPT-4o and human implementation. We tested the correctness of GPT-4o's SHA1 implementation by using the reference test vectors (predefined inputs with their expected outputs) provided by the National Software Reference Library (NSRL)~\cite{nsrl}.
GPT-4o's implementation produced {\em incorrect hash values for all the inputs}.
Such silent errors will not prompt any error message, but incorrect hash values can lead to security vulnerabilities and malfunctions in software that uses hash algorithm for authentication or integrity checks.
As expected, OpenBSD's SHA1 implementation computed all the hash values correctly.

\begin{figure*}[t]
\centering
\begin{tabular}{cc}
Human & GPT-4o \\
\begin{minipage}{.40\textwidth}
\begin{lstlisting}[{xleftmargin=1em}]
Graph newGraph(int V)
{
    assert(V >= 0);
    Graph g = malloc(sizeof(GraphRep));
    assert(g != NULL);
\end{lstlisting}
\end{minipage}
&
\hspace*{9mm}
\begin{minipage}{.5\textwidth}
\begin{lstlisting}
Graph newGraph(int V) // V can be negative!
{   // not checked for NULL
    Graph g = malloc(sizeof(GraphRep));
    // not checked for NULL
    g->edges = malloc(V * sizeof(int *));
\end{lstlisting}
\end{minipage}
\end{tabular}
\caption{LLM-generated code for a graph implementation, shown on the right, has no guardrails (comments added by us).}
\label{fig:graph_algo}
\end{figure*}

\subsection{Sample Task 3: Implementing Graph Data Structure}
\label{sec:graph_sample}

This task entailed generating a simple graph implementation, with functions for creating (allocating) a graph, adding and removing edges, printing the graph, and freeing (deallocating) the graph memory from the heap.
The relevant allocation code -- creating a graph of size \ls|V| -- is shown in \Cref{fig:graph_algo}. Note that the human-developed code, shown on the left, has checks for the value of \ls|V|, whereas the LLM-generated code, on the right, does not contain such checks. The LLM code has two issues. First, it does not check the result of \ls|malloc()| for NULL, hence opening the
first entry
for potential NULL pointer dereferences if the memory allocation fails.
Second, the LLM-generated code does not check for a potential negative value of \ls|V|. If a negative \ls|V| is passed as an argument,
the call to \ls|malloc()| on line~5
will most likely result
in a \ls|NULL| pointer (since \ls|malloc()| takes an unsigned argument, the argument is interpreted as a positive value
at the size of \ls|size_t_max| \ls|-| \ls|abs(V)|  \ls|+| \ls|1|, where \ls|size_t_max| is the max value of the type),
or, less likely, in a very large memory allocation. Therefore, memory allocation either fails or its size is in the control of a potential adversary.
We refer to this as \textit{malloc overflow}, a term used by Clang analyzer.

\subsection{Motivations}
Aligned with the national goal of managing the risk of generative AI~\cite{nist_ai_risk, ai_executive_order}, our research investigates the security of LLM-generated code.
Our research is motivated by the following  questions:

\noindent\textbf{Research Question 1:} Does LLM generate insecure code?

\noindent\textbf{Research Question 2:} For the same task, does LLM generate code that is more secure (or higher quality) than the human-written version?

\noindent\textbf{Research Question 3:} Can prompt engineering improve the security of LLM-generated code?

\section{Methodology}
\label{sec:method}

We select two categories of coding tasks to evaluate the security of LLM-generated code: (1) LeetCode problems, and (2) data structures and algorithms, including three widely-used cryptographic algorithms.
We chose these categories because they are widely used in almost all software engineering interviews and actual coding tasks, and they have relatively clear input-output interfaces.

For LeetCode problems, we selected all 202 problems from a well-known LeetCode solution repository on GitHub with 2.3k forks and 5.5k stars~\cite{neetcode_gh}.
We use the same set of data structures, algorithms, and LeetCode problems for both LLM-generated and human-written code to ensure a fair comparison.

For data structures and algorithms, we selected four sources of human code from GitHub: a collection of 13 data structures and algorithms in C, which has 4.3k forks and 18.8k stars~\cite{the_algorithms}; a repo with an additional data structure,  having 719 forks and 3.3k stars~\cite{fragglet}; a hashmap algorithm repo, with 205 forks and 520 stars~\cite{petewarden}; and the OpenBSD project for three cryptographic algorithms with 858 forks and 3.2k stars~\cite{openbsd}.

\begin{figure}[h!]
\centering
\includegraphics[width=0.44\textwidth]{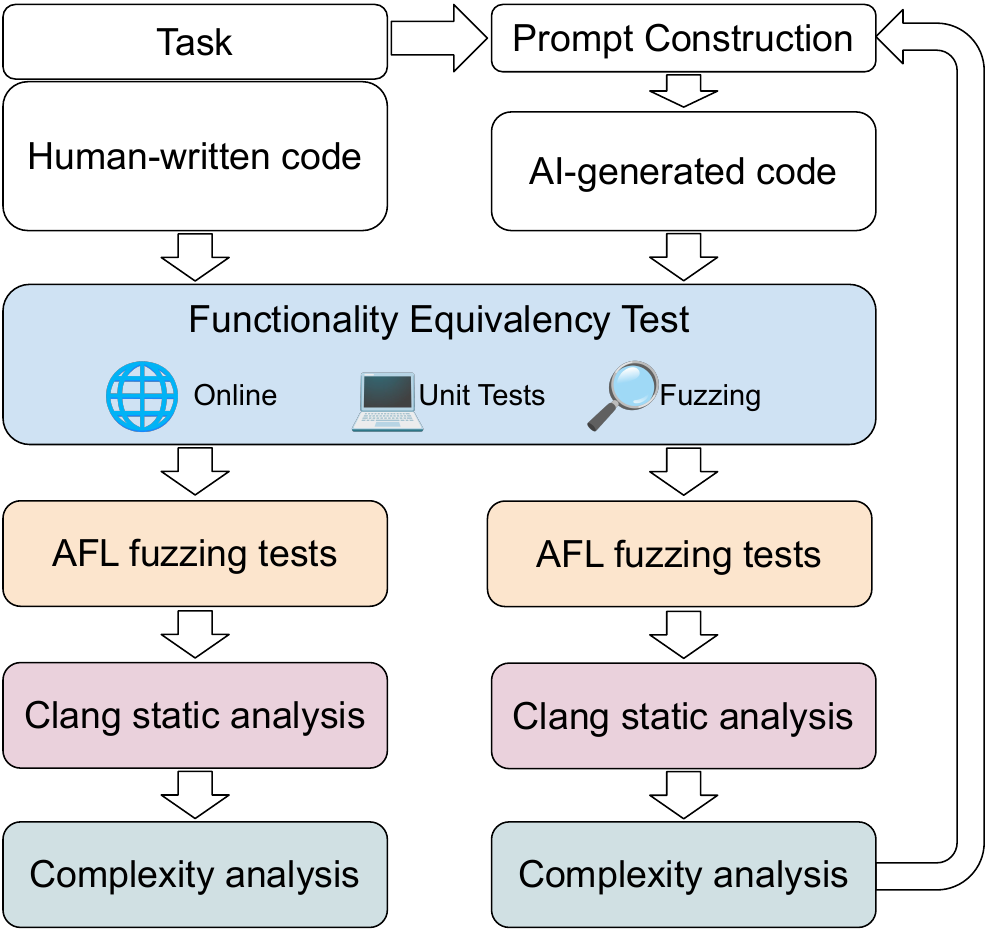}
\caption{The architecture of our framework, \Sname.}
\label{fig:arch}
\end{figure}

\subsection{LLM Code Generation}
\label{sec:llm_code_gen}

We use OpenAI's GPT-4o model to generate the full task solution for each coding task.
As shown in \Cref{fig:arch}, for each coding task, we provide its task description as the prompt to GPT-4o API, and collect the coding outputs.

For LeetCode problems, we locate the problem description in the LeetCode website using the problem title in the LeetCode solution repository, and use the description as a prompt.
The prompt is constructed in a way as close as possible to the actual task description given to a human developer.

For data structures and algorithms, we provided prompts for GPT-4o as follows: either the header files, or in the absence of header files, the function declarations (types) from the human-written code. This ensures that both GPT-4o and human implement the same set of functionalities and eliminate the discrepancies in function signatures
in code comparison.

\subsection{Functionality Validation}
\subsubsection{LeetCode Problems}
After obtaining the GPT-4o-generated code, we use test cases to validate the functionalities of the code against the human-written code for the same task.
If test cases are available, such as LeetCode online submission \cite{leetcode}, we submit GPT-4o-generated code directly to the LeetCode platform (\Cref{sec:leetcode_func_result}) for evaluation.

\subsubsection{Data Structures and Algorithms}
\label{sec:dsa_func_validation}

For data structures and algorithms, we employ unit testing and fuzzing to validate functionality, as follows.

{\bf Unit testing.} Test cases are manually written for unit testing with the help of CUnit~\cite{cunit}.
We examine the human-written code to understand the functionality provided by the data structure or algorithm, such as insertion, deletion, sorting, etc. We then write test cases, divided into 2 categories: (1) regular cases, and (2) edge cases.
Regular cases are typical scenarios that a function is expected to handle, such as inserting positive integers into a linked list, performing a merge sort on an unsorted array, etc.
Edge cases are crafted in a way to test a function's protection against unexpected inputs,
such as dequeuing from an empty queue, performing a breadth-first search on an empty graph,
sorting an empty array, etc.
We ran the same test cases on both GPT-4o-generated and human-written code, since we prompted GPT-4o to implement the data structures and algorithms with the same header files or function declarations as used in the human-written code.
By testing GPT-4o-generated and human-written code with the same test cases, we verify that they are achieving the same functionalities.

{\bf Fuzzing.} In addition to unit testing, we use AFL~\cite{afl} to fuzz both GPT-4o-generated code and human-written code.
We create an entry point to fuzz each data structure and algorithm. This entry point is the main method in each program where it reads an input file that contains various instructions and input values. For example, the line \ls|insert 10| in an input file will carry out the insert operation in the binary search tree and insert the value \ls|10| into the tree. The input file is used as a seed in fuzzing. Each data structure and algorithm has its own set of instructions to make sure that all implemented functionalities will be tested. Since we only focus on functionality validation, any invalid instructions mutated by AFL~\cite{afl} in the process of fuzzing will be ignored at the entry point. However, values will be mutated during fuzzing and GPT-4o-generated code and human-written code need to handle any kinds of invalid inputs. As it was not relevant to the comparison, the entry point code was not included in the security and complexity analysis.

\subsubsection{Cryptographic Algorithms}
\label{sec:crypto_func_validation}
For cryptographic algorithms, we employ a different way of validating the functionality.
For one-way hashing algorithms such as SHA1 and Message Digest Algorithm 5 (MD5), we encrypt the predefined inputs from test vectors and compare the computed hash values with the expected hash values from the test vectors; we used existing sets of test vectors (strings) originating from NSRL~\cite{nsrl} (as discussed in \Cref{sec:sha1_sample}).
For two-way
algorithms such as Advanced Encryption Standard (AES), we encrypt the predefined inputs and decrypt the encrypted values; we then validate the decrypted values against the original inputs.

\subsection{Static Analysis}
\label{sec:static_analysis}

\subsubsection{Security}

We use the Clang static analyzer~\cite{clang} to perform static analysis on both GPT-4o-generated and human-written code.
Clang provides a set of default checkers, such as null dereference, memory leak, and null arguments.
In addition, Clang provides a range of experimental (advanced) checkers in several categories, as follows. ``Core'' experimental checkers include detectors for pointer arithmetic, invalid casting/conversion, etc.\@ ``Security'' experimental checkers look for errors such as array index out of bounds, malloc overflow, etc.\@ ``Unix'' experimental checkers look for issues such as memory leaks or null pointers passed to string functions.
While the experimental checkers might, in theory, emit a higher number of false positives, our manual validation of the reported errors indicates a negligibly low rate of false positives.

\subsubsection{Complexity}

We  evaluate code complexity using several metrics: cyclomatic complexity, normalized complexity, and lines of code. Cyclomatic complexity is used to measure the complexity of a program's control flow~\cite{cyclcomp}.
High cyclomatic complexity can indicate that the code is harder to understand and maintain, and potentially prone to errors. As cyclomatic complexity is an absolute value that characterizes an entire file, larger files naturally have higher complexity. Therefore, to gauge the complexity of typical code in a file, we also compute the normalized complexity, i.e., divide complexity by the number of lines of code in that file -- this indicates the typical complexity expected for a code snippet.
We use SCC~\cite{scc} to obtain the cyclomatic complexity and the lines of code in each C file.
By ``lines of code'' we mean actual source code, excluding comments and blank lines.

\section{Analysis Results}
\label{sec:result}

We now present our comparative analysis of GPT-4o-generated and human-written code in terms of functionality (\Cref{sec:func_result}),  security
(\Cref{sec:sec_result}), and complexity/code quality (\Cref{sec:complexity_result}).

\subsection{Functionality Analysis}
\label{sec:func_result}

\subsubsection{LeetCode Problems}
\label{sec:leetcode_func_result}

The LeetCode platform provides an online submission system \cite{leetcode} where submitted code has to pass all the test cases -- only then it is considered a correct solution. All the human-written code that we gathered from the GitHub repository \cite{neetcode_gh} had to pass all the test cases from LeetCode online submission system prior to being placed in the repository. Therefore, we only tested GPT-4o-generated code through LeetCode online submission~\cite{leetcode}. We found that 87.6\% of  LeetCode solutions generated by GPT-4o passed the LeetCode online submission checks.
The 12.4\% of LeetCode solutions generated by GPT-4o that failed were due to issues such as exceeding time limit, failing test cases, and generating runtime or compile time errors. The detailed analysis results are reported in \Cref{tab:leetcode_results} and discussed next.

{\bf Failing test cases and exceeding time limit.}
Among the solutions with issues, 9 (39.6\%) are failing test cases and exceeding time limit. Although detailed instructions with sample inputs and outputs are given in the prompts,
GPT-4o's solutions are not always able to fulfill all the requirements stated in the prompts.

{\bf Runtime errors.}
In addition to functionality test cases, LeetCode's online checker compiles our submitted code with Address Sanitizer enabled, and looks for runtime errors as well, such as memory violations and other crashes.
Runtime errors take up a significant 48\% of the solutions with issues. There are multiple categories of runtime errors such as buffer overflow, signed integer overflow, array index out of bounds, and load of address with insufficient space.
The results highlight the importance of dynamic (runtime analysis) in conjunction with static analysis: 8 out of 12 of the runtime errors were not identified during our Clang static analysis. Consequently, these silent, though critical errors, could easily be neglected by users and introduce severe security threats if this code were to be used in real-world programs -- a scenario that is very likely to happen, as pointed out in prior research and mentioned in \Cref{sec:intro}.

{\bf Compile time errors.} Compile time errors are due to incorrect function declarations generated by GPT-4o. These issues can be mitigated by providing specific function declarations required by LeetCode in the prompts.

\begin{table*}[h]
\centering
\begin{tabular}{|c|c|c|p{9.5cm}|}
\hline
\textbf{Question\#} & \textbf{Error Type} & \textbf{Difficulty} & \textbf{Error Details} \\
\hline
215  & Time Limit Exceeded & Medium & \\
\hline
853  & Time Limit Exceeded & Medium & \\
\hline
115  & Runtime Error       & Hard   & signed integer overflow \\
\hline
297  & Runtime Error       & Hard   & buffer overflow \\
\hline
332  & Runtime Error       & Hard   & load of address with insufficient space \\
\hline
22   & Runtime Error       & Medium & heap-buffer-overflow \\
\hline
152  & Runtime Error       & Medium & signed integer overflow \\
\hline
179  & Runtime Error       & Medium & stack buffer overflow \\
\hline
377  & Runtime Error       & Medium & signed integer overflow \\
\hline
523  & Runtime Error       & Medium & array index out of bounds \\
\hline
846  & Runtime Error       & Medium & load of address with insufficient space \\
\hline
981  & Runtime Error       & Medium & AddressSanitizer: requested allocation size exceeds maximum supported size \\
\hline
1838 & Runtime Error       & Medium & signed integer overflow \\
\hline
208  & Runtime Error       & Medium & heap-buffer-overflow \\
\hline
1046 & Failed Test Case    & Easy   & \\
\hline
51   & Failed Test Case    & Hard   & \\
\hline
212  & Failed Test Case    & Hard   & \\
\hline
40   & Failed Test Case    & Medium & \\
\hline
138  & Failed Test Case    & Medium & \\
\hline
435  & Failed Test Case    & Medium & \\
\hline
1930 & Failed Test Case    & Medium & \\
\hline
88   & Compile Error       & Easy   & function definition - number of parameters invalid \\
\hline
1299 & Compile Error       & Easy   & function definition - number of parameters invalid \\
\hline
1899 & Compile Error       & Medium & function definition - number of parameters invalid \\
\hline
1968 & Compile Error       & Medium & function definition - number of parameters invalid \\
\hline
\end{tabular}
\caption{LeetCode-reported issues in GPT-4o-generated code.}
\label{tab:leetcode_results}
\end{table*}

\subsubsection{Data Structures and Algorithms}

\Cref{sec:dsa_func_validation} discussed our strategy for writing unit test cases and seed files, used in unit testing and fuzzing, respectively. We discuss our findings next.

{\bf Unit testing.} All of the data structures and algorithms implemented by GPT-4o passed the unit testing. For human-written code, 3 out of 15 of the data structures and algorithms did not pass unit testing due to failing test cases and causing segmentation faults. For example, in the implementation of doubly-linked list, the usage of \ls|==| for double comparison leads to failing test cases. In the implementation of red-black trees,
a missed \ls|return| statement causes \ls|nullptr| to be used as a buffer address, leading to null pointer dereference.

{\bf Fuzzing.} In terms of unique hangs, GPT-4o-generated code has 50\% more hangs than human-written code, whereas for unique crashes, GPT-4o generated code has 23.9\% more crashes than human-written code. We were unable to perform fuzzing on certain human-written code such as queue, doubly linked list, and red black tree due to the bug mentioned above in this section (the null pointer dereference crash caused by the missed \ls|return| statement).
These files are listed as N/A in \Cref{tab:fuzzing_results}. Note that these unit tests focused on functionality alone; the security aspect will be discussed in \Cref{sec:sec_result}.

\begin{table}[h!]
\centering
\begin{tabular}{|p{2.15cm}|c|c|c|c|}
\hline
\multirow{2}{*}{} & \multicolumn{2}{c|}{\textbf{Unique Hangs}} & \multicolumn{2}{c|}{\textbf{Unique Crashes}} \\
\cline{2-5}
& \textbf{LLM}  & \textbf{Human}    & \textbf{LLM}  & \textbf{Human}\\ \hline
Stack               & 3             & 4                 & 0             & 0             \\ \hline
Queue               & 0             & N/A               & 13            & N/A           \\ \hline
Singly Linked List  & 0             & 0                 & 65            & 73            \\ \hline
Doubly Linked List  & 1             & N/A               & 59            & N/A           \\ \hline
Binary Search Tree  & 0             & 0                 & 48            & 65            \\ \hline
AVL Tree            & 0             & 0                 & 58            & 64            \\ \hline
Red Black Tree      & 0             & N/A               & 45            & N/A           \\ \hline
Hash Map            & 0             & 0                 & 16            & 27            \\ \hline
Hash Set            & 0             & 0                 & 22            & 15            \\ \hline
Hash Table          & 0             & 0                 & 19            & 15            \\ \hline
Graph               & 2             & 0                 & 25            & 40            \\ \hline
Breadth First Search& 3             & 0                 & 27            & 33            \\ \hline
Depth First Search  & 1             & 0                 & 54            & 32            \\ \hline
Merge Sort          & 1             & 1                 & 0             & 0             \\ \hline
Bubble Sort         & 1             & 1                 & 0             & 0             \\ \hline
\textbf{Total}      & 12     &\cellcolor{green!30}6     & 451     &\cellcolor{green!30}364 \\ \hline
\end{tabular}
\caption{Unique hangs and crashes from AFL fuzzing on GPT-4o and human-written code. Green means lower in values.}
\label{tab:fuzzing_results}
\end{table}

\subsubsection{Cryptographic Algorithms}

We performed unit testing on AES, MD5, and SHA1 as discussed in \Cref{sec:crypto_func_validation}.
Human-written code from OpenBSD generates all the correct hash values, as anticipated.
GPT-4o's implementations of MD5 and AES compute all values correctly.
GPT-4o's SHA1 implementation computes {\em wrong hash values for all the inputs}, which could be catastrophic if this implementation were to be used in real-world situations.
Functionality issues like this have a high chance of going unnoticed, unless the LLM-generated code goes through strict testing.

\subsection{Security Analysis}
\label{sec:sec_result}

We perform static analysis on both GPT-4o-generated and human-written code using Clang static analyzer~\cite{clang}.
\Cref{tab:sec_analysis_results} shows the analysis results of 220 files (202 LeetCode problems, 18 data structures and algorithms).
Overall, GPT-4o-generated code has 10.3\% more security issues than human-written code. The three categories of issues that have relatively higher counts are \ls|malloc| overflow, array index out of bounds, and memory leak. We found that GPT-4o-generated code has 32.8\% more \ls|malloc| overflow issues than human-written code.
This is a consequence of GPT-4o's tendency of assuming all inputs are valid (\Cref{fig:graph_algo}). The counts for array index out of bounds issues are similar for both GPT-4o and human. \Cref{fig:leetcode_array_index_out_of_bounds_0010} illustrates the similar mistakes made by both GPT-4o and human in one of the LeetCode problems where both code may access elements that are out of bounds. Memory leak issues, where human code has a 50\% higher count than GPT-4o, are mainly due to the problem description, where the LeetCode problems specifically mention that the caller will invoke \ls|free()|.
However, the memory leak issues do exist if the code snippet is used outside the context of these LeetCode questions.

Other than issues reported by Clang, we also look at the implementation details between GPT-4o-generated and human-written code on data structures and algorithms. We do not focus on the implementation details of LeetCode solutions because the main objective of a LeetCode problem is to provide a solution that solves a very particular problem with a given set of constraints that passes all the test cases provided.

We found that GPT-4o has a tendency of generating exact same code as human-written code. We will refer to this as ``code parroting'' for the rest of the paper. Note that we do not include any implementation details of human-written code in the prompts with the exception of header files or function declarations. In GPT-4o-generated code, 2 out of the 18 files in data structures and algorithms are the exact replica of human-written code, as follows. In AVL tree's implementation, GPT-4o reproduces the human-written code from GitHub, including the helper functions. Similarly, in the graph data structure implementation, GPT-4o reproduces the human-written code, but fails to add validation for input values (\Cref{fig:graph_algo}). Code parroting is extremely harmful considering the fact that malicious code could possibly be the training data of LLM (more commonly known as data poisoning if malicious information is injected into training data on purpose). Harmful code will then be generated by LLM and possibly utilized by developers.

\begin{table}[h!]
\centering
\begin{tabular}{|p{4.9cm}|c|c|}
\hline
\textbf{Category}                               & \textbf{GPT-4o}   & \textbf{Human} \\ \hline
Memory Leak                                     & 8                 & 16    \\ \hline
Null pointer dereference                        & 6                 & 8     \\ \hline
Assigned value is garbage or undefined          & 2                 & 2     \\ \hline
Malloc overflow                                 & 81                & 61    \\ \hline
Taint propagation                               & 0                 & 1     \\ \hline
Division by zero                                & 0                 & 1     \\ \hline
Address of stack memory returned to caller      & 0                 & 1     \\ \hline
Undefined binary operator                       & 3                 & 3     \\ \hline
Result of \ls|malloc| converted to invalid type & 0                 & 1     \\ \hline
Undefined or garbage value returned to caller   & 3                 & 1     \\ \hline
Function call argument is an uninitialized value& 2                 & 3     \\ \hline
Array index out of bounds                       & 16                & 17    \\ \hline
Implicit conversion                             & 4                 & 1     \\ \hline
Nested function is not supported                & 1                 & 0     \\ \hline
Cast from non struct to struct                  & 1                 & 0     \\ \hline
C string out of bounds                          & 1                 & 0     \\ \hline
\textbf{Total}                                  & 128               & \cellcolor{green!30}116   \\ \hline
\end{tabular}
\caption{Number of issues found by the Clang static analyzer on GPT-4o and human-written code. }
\label{tab:sec_analysis_results}
\end{table}

\begin{figure*}[t]
\centering
\begin{tabular}{cc}
Human & GPT-4o \\
\begin{minipage}{.45\textwidth}
\begin{lstlisting}[{xleftmargin=1em}]
// j starts from 1
for (int j=1; j<=n; j++) {
    if (p[j-1] == s[i-1] || p[j-1] == '.') {
        dp[i][j] = dp[i-1][j-1];
    }
    else if (p[j - 1] == '*') {
        // [j-2] could be -1
        dp[i][j] = dp[i][j-2] || ((s[i - 1] == p[j-2] || p[j-2] == '.') && dp[i-1][j]);
\end{lstlisting}
\end{minipage}
&
\hspace*{7mm}
\begin{minipage}{.45\textwidth}
\begin{lstlisting}[{xleftmargin=1em}]
for (int j=1; j<=n; j++) { // j starts from 1
    if (p[j-1] != '*') {
        dp[i][j] = i > 0 && dp[i-1][j-1] && (s[i-1] == p[j-1] || p[j-1] == '.');
    } else {
        // [j-2] could be -1
        dp[i][j] = (j>=2 && dp[i][j-2]) || (i>0 && dp[i-1][j] && (s[i-1] == p[j-2] || p[j-2] == '.'));
\end{lstlisting}
\end{minipage}
\end{tabular}
\caption{Array index out of bounds issue.}
\label{fig:leetcode_array_index_out_of_bounds_0010}
\end{figure*}

\subsection{Complexity Analysis}
\label{sec:complexity_result}
SCC \cite{scc} is the tool that we utilize to measure cyclomatic complexity and lines of code (LoC, excluding comments and blank lines) per file.
We calculate the mean, median, geometric mean, and trimmed mean for both complexity and normalized complexity (complexity per LoC).

As shown in \Cref{tab:scc_complexity_results}, GPT-4o-generated code seems to have lower complexity on the surface. However, it would be incorrect to assume that this lower complexity indicates high-quality code. First, low (absolute) complexity is an artifact
of GPT-4o generating less code than the human equivalent for the same task. In fact, human-written code always has a lower {\it normalized complexity}, which is a more indicative metric of the efforts needed to understand and maintain a typical snippet of code in a given file. Second, the low complexity is due to GPT-4o generating ``bare-bones'' code. For instance, the GPT-4o-generated hash map implementation (\Cref{fig:gpt_hash_map}) uses a simple hash function that does not check for collisions, while the human-written implementation has a more sophisticated hash function that checks for collisions (\Cref{fig:human_hash_map}). This example illustrates one of the reasons why the GPT-4o-generated code has lower cyclomatic complexity (when computing complexity for the entire file, i.e., without normalizing for lines of code). Another example is in \Cref{fig:leetcode_input_verification_0981} where the human-written implementation validates input values before carrying out certain operations in one of the LeetCode problems, whereas GPT-4o makes the assumptions that all inputs are valid. This undoubtedly increases the cyclomatic complexity of human-written code but illustrates that human-written code uses {\it defensive programming}, which in turn mitigates various security threats.

\begin{figure}[h!]
\centering
\includegraphics[width=0.51\textwidth, trim={0 1cm 0 1cm}]{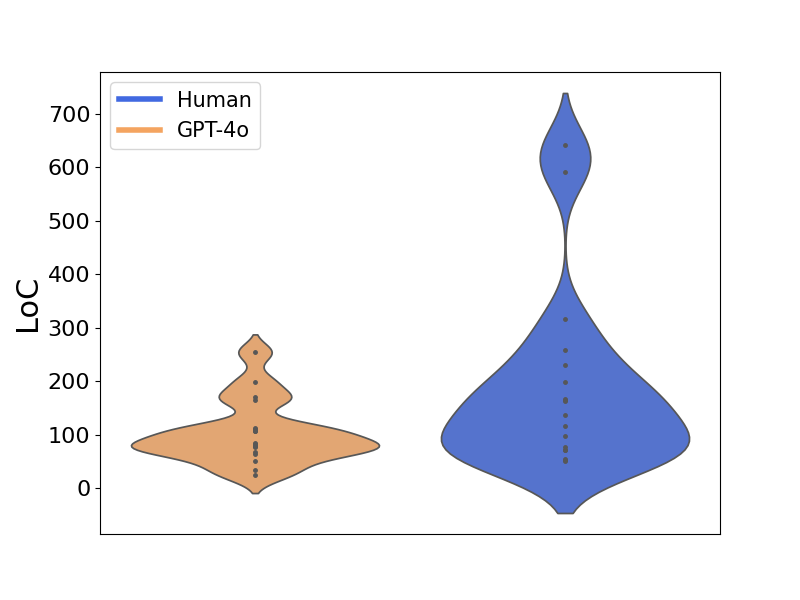}
\caption{LoC per file in data structures and algorithms.}
\label{fig:loc_ds_algo}
\end{figure}

\begin{figure}[h!]
\centering
\includegraphics[width=0.51\textwidth, trim={0 1cm 0 1cm}]{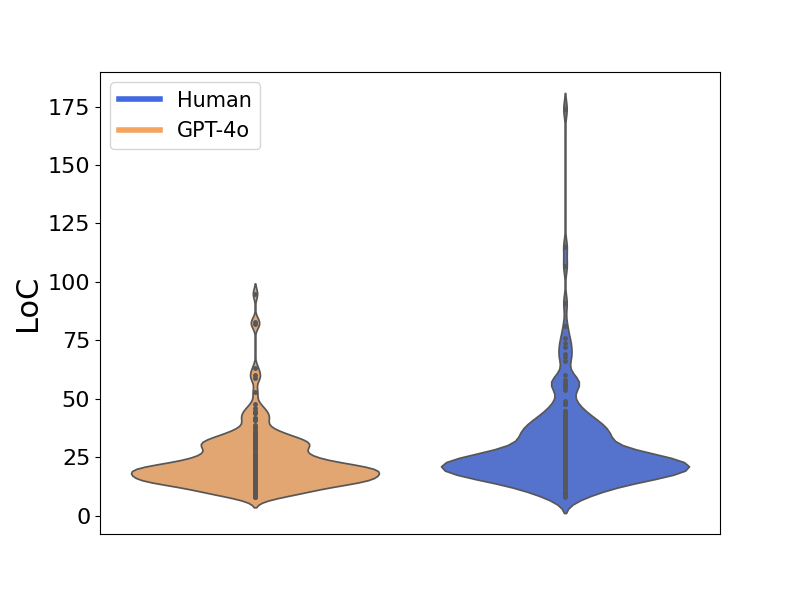}
\caption{LoC per file in LeetCode.}
\label{fig:loc_leetcode}
\end{figure}

In terms of LoC, the results in \Cref{tab:scc_loc_results}, \Cref{fig:loc_ds_algo}, and \Cref{fig:loc_leetcode} demonstrate that GPT-4o always produces fewer LoC than human. However, this is due to the lack of defensive programming as shown in \Cref{fig:graph_algo,fig:leetcode_input_verification_0981} and simplification of certain functionalities as shown in \Cref{fig:gpt_hash_map} in LLM-generated code.
Undoubtedly, GPT-4o-generated code being more concise than human-written code could serve a valuable purpose, e.g., in introductory programming classes.
That being the case, introducing security risks into real-world programs seems to be unavoidable if LLM-generated code is utilized in a professional environment.

\begin{table*}[h!]
\centering
\begin{tabular}{|c|c|c|c|c|c|c|c|c|c|c|c|c|c|c|c|c|}
\hline
\multirow{2}{*}{} & \multicolumn{4}{c|}{\textbf{Complexity (LC)}} & \multicolumn{4}{c|}{\textbf{Complexity/Code (LC)}} & \multicolumn{4}{c|}{\textbf{Complexity (D\&A)}} & \multicolumn{4}{c|}{\textbf{Complexity/Code (D\&A)}} \\ \cline{2-17}
& \textbf{Mn} & \textbf{Md} & \textbf{GM} & \textbf{TM} & \textbf{Mn} & \textbf{Md} & \textbf{GM} & \textbf{TM}
& \textbf{Mn} & \textbf{Md} & \textbf{GM} & \textbf{TM} & \textbf{Mn} & \textbf{Md} & \textbf{GM} & \textbf{TM} \\ \hline
\!\!GPT-4o\!\!
& \cellcolor{green!30}7.37    & 7    & \cellcolor{green!30}6.03        & \cellcolor{green!30}7.07
& 0.33        & 0.29        & 0.31        & 0.21
& \cellcolor{green!30}25.7    & \cellcolor{green!30}14      & \cellcolor{green!30}16.4        & \cellcolor{green!30}18.8
& 0.21        & 0.23        & 0.18        & 0.21
\\ \hline
\!\!Human\!\!
& 8.10        & 7           & 6.19        & 7.44
& \cellcolor{green!30}0.28  & \cellcolor{green!30}0.25   & \cellcolor{green!30}0.27    & \cellcolor{green!30}0.16
& 34.5        & 23          & 21.2        & 25.1
& \cellcolor{green!30}0.17  & \cellcolor{green!30}0.16   & \cellcolor{green!30}0.15    & \cellcolor{green!30}0.16

\\ \hline
\end{tabular}
\caption{Cyclomatic complexity measurements. LC is LeetCode (202 files), D\&A is data structures and algorithms (18 files), Mn is mean, Md is median, GM is geometric mean, and TM is trimmed mean. We use the same abbreviations for the rest of the paper. Trimmed mean is at 5\% for LeetCode and 15\% for data structures and algorithms.}
\label{tab:scc_complexity_results}
\end{table*}

\begin{table*}[h!]
\centering
\begin{tabular}{|c|c|c|c|c|c|c|c|c|}
\hline
\multirow{2}{*}{} & \multicolumn{4}{c|}{\textbf{Lines of Code (LC)}} & \multicolumn{4}{c|}{\textbf{Lines of Code (D\&A)}} \\ \cline{2-9}
& \textbf{Mn} & \textbf{Md} & \textbf{GM} & \textbf{TM} & \textbf{Mn} & \textbf{Md} & \textbf{GM} & \textbf{TM} \\ \hline
GPT-4o
& \cellcolor{green!30}23.6        & \cellcolor{green!30}20        & \cellcolor{green!30}21.2        & \cellcolor{green!30}22.2  & \cellcolor{green!30}103.5       & \cellcolor{green!30}83        & \cellcolor{green!30}88.7        & \cellcolor{green!30}99.1 \\ \hline
Human   & 29.6        & 24        & 25.6        & 27.2        & 192.1        & 149       & 142         & 172.8\\ \hline
\end{tabular}
\caption{Lines of Code measurements. Trimmed mean is at 5\% for LeetCode and 10\% for data structures and algorithms.}
\label{tab:scc_loc_results}
\end{table*}

\begin{figure}[t]
\centering
\begin{minipage}{.45\textwidth}
\begin{lstlisting}[{xleftmargin=1em}]
unsigned long hashFunction(const char *key) {
    unsigned long hash = 0;
    while (*key) {
        hash = (hash << 5) + *key++;
    }
    return hash;
}
\end{lstlisting}
\end{minipage}
\caption{GPT-4o implementation of hashing in Hash Map.}
\label{fig:gpt_hash_map}
\end{figure}

\begin{figure}[t]
\centering
\begin{minipage}{.45\textwidth}
\begin{lstlisting}[{xleftmargin=1em}]
int hashmap_hash(map_t in, char* key){
	int curr;
	int i;

	/* Cast the hashmap */
	hashmap_map* m = (hashmap_map *) in;

	/* If full, return immediately */
	if(m->size >= (m->table_size/2))
            return MAP_FULL;

	/* Find the best index */
	curr = hashmap_hash_int(m, key);

	/* Linear probing */
	for(i = 0; i< MAX_CHAIN_LENGTH; i++){
		if(m->data[curr].in_use == 0)
			return curr;

		if(m->data[curr].in_use==1 && (strcmp(m->data[curr].key,key)==0))
			return curr;

		curr = (curr + 1) % m->table_size;
	}
	return MAP_FULL;
}
\end{lstlisting}
\end{minipage}
\caption{Human implementation of hashing in Hash Map.}
\label{fig:human_hash_map}
\end{figure}

\begin{figure*}[t]
\centering
\begin{tabular}{cc}
Human & GPT-4o \\
\begin{minipage}{.45\textwidth}
\begin{lstlisting}[{xleftmargin=1em}]
void timeMapSet(TimeMap* obj, char* key, char* value, int timestamp) {

    if (obj == NULL || key == NULL || value == NULL)
        return;

    int index = -1;
    for (int i = 0; i < obj->size; i++) {
\end{lstlisting}
\end{minipage}
&
\hspace*{7mm}
\begin{minipage}{.45\textwidth}
\begin{lstlisting}[{xleftmargin=1em}]
void timeMapSet(TimeMap* obj, char* key, char* value, int timestamp) {

    // no input verifications
    for (int i = 0; i < obj->node_count; i++) {
        obj->nodes[i].entries[obj->nodes[i].entry_count].timestamp = timestamp;
\end{lstlisting}
\end{minipage}
\end{tabular}
\caption{GPT-4o-generated code (right) does not verify inputs.}
\label{fig:leetcode_input_verification_0981}
\end{figure*}

\section{Feedback Loop}
\label{sec:loop_test_results}

\subsection{Rationale and Setup}

State-of-the-art practice suggests using a reinforcement loop (\Cref{sec:ar_transformer}) to improve code quality when using LLMs. Therefore, we constructed a feedback loop as a way to improve GPT-4o-generated code, starting by trying to eliminate the security issues identified in \Cref{sec:sec_result}. For this experiment, we included 60 LeetCode files (solutions) generated by GPT-4o, as follows: 30 files with security issues (found by Clang), and 30 files without security issues.
The list of security issues is listed in \Cref{tab:loop_sec_issues}. We ask GPT-4o to regenerate these files based on different security categories.
In each loop, prompts for files with security issues will be accompanied by prompts for files that have no security issues.
For instance, we specifically ask GPT-4o to fix only malloc overflow issues in this loop with a total of 18 prompts related to files that have malloc overflow issues along with another 18 prompts related to files that have no malloc overflow issues. Note that the prompts (LeetCode problem descriptions) used in the loops are the same as the non-feedback setup, but with additional instructions that ask GPT-4o to make sure there are no certain security issues in the code.
Since there are 3 categories of security issues, a total of 3 loops will be conducted, each loop to fix a specific security issue.

\subsection{Results}

{\bf Files with security issues.}
For files that did contain security issues, GPT-4o managed to lower the number of security issues found after one iteration on malloc overflow and null dereference from 40 to 24 and 3 to 2, respectively, as shown in \Cref{tab:loop_sec_1}. However, {\it the security issues are not completely removed, even though we specifically asked GPT-4o to ensure the code will not have these security issues}. Furthermore, for array index out of bounds issues, GPT-4o {\it generated 3 more such issues}. This is a critical finding since it shows the possibility that LLM may generate code that is the exact opposite of what users have specified in the prompt.

{\bf Files without security issues.}
For files without any security issue before entering the loop, GPT-4o introduces 4 new malloc overflow issues as shown in \Cref{tab:loop_sec_2}.
\Cref{fig:loop_security_before_after} illustrates one of the code examples (before and after running the loop), where the new code has a malloc overflow issue.
Malloc overflows account for 63.3\% of the total security issues found across the code generated by GPT-4o. However, GPT-4o-generated code, on the right side of \Cref{fig:loop_security_before_after}, still does not check the value of \ls|k| and the result of \ls|malloc| after an effort has been made in the prompt to ask LLM to make no such mistakes. This outcome is especially troublesome, since the previously-produced solutions contained no such issues.
Note that the particular solution on the left in \Cref{fig:loop_security_before_after} had no prior issues found during static analysis but did have a runtime error (array index out of bounds) when submitted to the LeetCode online submission platform \cite{leetcode} before running the feedback loop. However, the regenerated solution, supposedly the ``better'' solution, on the right in \Cref{fig:loop_security_before_after}, still has a runtime error, i.e., exceeding memory limit when \ls|k = 1000000000| (exposed by LeetCode's online submission system).
As LeetCode runtime checker caught this bug as well, it validates the findings from Clang analyzer.

\subsection{Why can't Prompt Engineering Fix Security Issues?}

The results from the feedback loop allow us to draw a conclusion that LLM so far is not capable of consistently removing security issues even though the prompt has explicitly asked it to do so. LLM might even go against the prompt in producing more security issues in the code.
First of all, these security issues are found by running static analysis. Users who utilize LLM for code generation might not even notice that the code poses serious security threats without running static analysis. Furthermore, even static analysis is not enough: some of the runtime errors found by LeetCode's online submission system~\cite{leetcode} were not identified during static analysis (\Cref{sec:leetcode_func_result}). To conclude, even for users who are aware of potential issues and diligent in attempting to reduce or eliminate security risks in the code, the LLM route might be unfruitful: LLM regeneration may worsen the security problem in the generated code, e.g., as shown in \Cref{fig:loop_security_before_after}.

\begin{table}[t]
\centering
\begin{tabular}{|c|c|c|c|c|}
\hline
& \textbf{M}    & \textbf{A}    & \textbf{N}    & \textbf{Total} \\ \hline
\# of files &  18           &10             & 2             & 30             \\ \hline
\end{tabular}
\caption{Number of files used in loop testing based on security categories. M is malloc overflow, A is array index out of bounds, N is null dereference. We use the same abbreviations for the rest of the paper.}
\label{tab:loop_sec_issues}
\end{table}

\begin{table}[t]
\centering
\begin{tabular}{|c|c|c|c|c|}
\hline
& \textbf{M}    & \textbf{A}    & \textbf{N}    & \textbf{Total} \\ \hline
Before      &  40           & 15            & 3             & 58             \\ \hline
After       &  24           & \cellcolor{red!30}18            & 2             & 44             \\ \hline
\end{tabular}
\caption{Number of security issues found before and after running the loop on files with security issues before based on security categories. Red means an increase in issues found.}
\label{tab:loop_sec_1}
\end{table}

\begin{table}[t]
\centering
\begin{tabular}{|c|c|c|c|c|}
\hline
& \textbf{M}    & \textbf{A}    & \textbf{N}    & \textbf{Total} \\ \hline
Before      &  0            & 0             & 0             & 0              \\ \hline
After       &  \cellcolor{red!30}4            & 0             & 0             & 4              \\ \hline
\end{tabular}
\caption{Number of security issues found before and after running the loop on files without security issues before based on security categories.}
\label{tab:loop_sec_2}
\end{table}

\begin{figure*}[t]
\centering
\begin{tabular}{cc}
Before & After \\
\begin{minipage}{.45\textwidth}
\begin{lstlisting}[{xleftmargin=1em}]
bool checkSubarraySum(int* nums, int numsSize, int k) {
    if (numsSize < 2)
        return false;

    int sum = 0;
    int remainder;
    int remainder_map[numsSize];

    for (int i = 0; i < numsSize; i++) {
        remainder_map[i] = -1;
    }

    remainder_map[0] = 0;

    for (int i = 0; i < numsSize; i++) {
        sum += nums[i];
        remainder = sum % k;

        if (remainder < 0) remainder += k;

        if (remainder_map[remainder] < 0) {
            remainder_map[remainder] = i + 1;
        } else if (i - remainder_map[remainder] > 0) {
            return true;
        }
    }
    return false;
}
\end{lstlisting}
\end{minipage}
&
\hspace*{7mm}
\centering
\begin{minipage}{.45\textwidth}
\begin{lstlisting}[{xleftmargin=1em}]
bool checkSubarraySum(int* nums, int numsSize, int k) {
    // not checking k before malloc
    int *remToIndex = (int *)malloc(k * sizeof(int));
    // not checking result of malloc
    for (int i = 0; i < k; i++) {
        remToIndex[i] = -1;
    }
    remToIndex[0] = 0;
    int cumulativeSum = 0;

    for (int i = 0; i < numsSize; i++) {
        cumulativeSum += nums[i];
        int remainder = cumulativeSum % k;
        if (remainder < 0) remainder += k;

        if (remToIndex[remainder] != -1) {
            if (i + 1 - remToIndex[remainder] >= 2) {
                free(remToIndex);
                return true;
            }
        } else {
            remToIndex[remainder] = i + 1;
        }
    }
    free(remToIndex);
    return false;
}
\end{lstlisting}
\end{minipage}
\end{tabular}
\caption{LeetCode solutions for problem number 523 generated by GPT-4o before and after running a loop.}
\label{fig:loop_security_before_after}
\end{figure*}

\section{Related Work}
\label{sec:related}

\subsection{AI-based Code Generation Security}
Similar studies have been done on the security aspect of LLM code generation. Perry et al. \cite{perry2023users} conducted a user study on two groups of participants: one without and the other with, access to an AI assistant. Participants were asked to solve a few specific tasks using multiple programming languages. Their results show that participants with access to the AI assistant wrote significantly less secure code, which aligns with our findings. We share a similar goal but take a different approach to addressing the AI assistant-introduced issues, as
the comparisons in our study are between LLM and human in solving algorithmic and data structure tasks.

We discuss  code parroting in \Cref{sec:sec_result}. Code parroting could be extremely dangerous when malicious code is injected into the training set (poisoning attack). In fact, numerous studies have shown that code completion tools that utilize LLM are vulnerable to poisoning attack such as suggesting insecure code in AES implementation \cite{schuster2021}. A tool, CodeBreaker~\cite{yan2024backdoorattack}, was designed for back-door attacks with the help of LLM where poisoned data for code generation can avoid strong vulnerability detection. These works support our point that LLM-generated code is insecure, requiring further analyses and improvements.

\subsection{Benchmarks for LLM-Generated Code}
Purple Llama CyberSecEval \cite{bhatt2023purplellama} and Cybench \cite{zhang2024cybench} are benchmarks  designed to tackle the security issues arising from LLM-generated code. CyberSecEval selects insecure code tests from real-world open source codebases whereas Cybench gathers 40 Capture the Flag (CTF) tasks for evaluating LLM-generated code.
These benchmarks are orthogonal to our purpose in evaluating an LLM's coding capabilities. \Sname is a comprehensive evaluation framework that can incorporate these benchmarks as a part of its testing if needed.

\subsection{Code Improvement in LLM}
As we discussed in \Cref{sec:ar_transformer}, RL has been employed by LLMs to improve code generation. Due to the implementation of these blackbox components, we would not know for sure if the improvements have been made in the security aspect of code generation. However, our study has shown that LLM-generated code produces a significant amount of security risks.
Prompt-based learning \cite{liu2021promptbasedlearning} is an approach where LLMs are guided to generate desired output with carefully designed prompts without re-training the models.
This technique could be applied to code generation as well. For instance, whenever \ls|malloc| appears in code generation, it should come with an allocation size check before using \ls|malloc| and a \ls|NULL| check for the result of \ls|malloc|. This would be one of the desired outputs to drastically reduce malloc overflow issues. However, based on the results we have seen so far in our study, LLM-generated code still has a long way to go to achieve this level of security.

\subsection{Privacy and Security in AI}

Code generation is not the only domain that raises privacy and security issues regarding LLMs.
Client-side prompt sanitization \cite{chong2024casper, yue2021differential,shen2024fire} is designed to protect private information shared with LLMs without affecting the performance of LLMs.
Federated learning \cite{mcmahan2017communication} can be used to train Machine Language models without sharing raw data that could be privacy-sensitive.
Prompt injection is an attack where LLMs could generate harmful content due to malicious prompts; benchmarks~\cite{liu2024promptinjectionbenchmark} and fine-tuned models~\cite{piet2024jatmopromptinjectiondefense} can defend against such attacks. Our study shows the importance of scrutinizing code security as well.

\section{Discussion}
\label{sec:discussion}

\subsection{Multiple LLMs}
Our study focuses on a single LLM, GPT-4o. We could extend our study by conducting the comparisons using the same human-written code with other LLMs such as Llama3, Claude, etc.; which would reveal differences between different LLMs in terms of code security.
However, the focal point of our study is to bring attention to the differences between LLM-generated and human-written code. We believe that the state-of-the-art LLM at the time of writing, GPT-4o, has successfully revealed the security threats brought upon by LLMs-based code generation; the issues we found are critical, yet likely overlooked when programming with the help of LLMs.

\subsection{Coding Tasks}
The coding tasks in our experiments are relatively small and independent as opposed to the code in large software projects such as an operating system kernel, a browser, etc.\@
Comparing code implemented for a subsystem can be an interesting future work since it requires more attention to interoperability, as these software has a plethora of components within themselves.
Another possible future work in terms of coding tasks could be asking LLM to fix known errors or identify possible issues in human-written code.

\section{Conclusions}
\label{sec:conclusion}

As AI-assisted code writing is spreading, it is imperative to study its impact and consequences on software security.
We developed a methodology and framework, \Sname, for comparing the security and quality
of LLM-generated and human-written code for the same task. We expose and
quantify the disadvantages of na\"ively using LLM-generated code in a critical domain such as security:
the LLM-generated code might violate the expected functionality in subtle ways, or might contain security issues that do not manifest until
late in program execution, and can be exploited by adversaries. In contrast, human code for equivalent tasks tends to contain guardrails and defensive constructs.
We also show the potential perils of using a feedback loop when programming with the help of LLMs.
Our work sheds light on the importance of scrutinizing LLM-generated code (and code re-generated by prompting) for functionality,  security, and quality issues. As long as LLM-generated code is prone to security risks it should be considered potentially harmful.

\section*{Acknowledgments}

This research was supported in part by the National Science Foundation grant CCF-2106710.
The authors also thank OpenAI for donating API credits to this project.

\balance
\bibliographystyle{plain}
\bibliography{zephyr}

\end{document}